# A GENERAL EXPRESSION FOR THE QUARTIC LOVELOCK TENSOR

C. C. Briggs

*Center for Academic Computing, Penn State University, University Park, PA 16802*

Tuesday, March 25, 1997, 7:37:49 PM

**Abstract.** A general expression is given for the quartic Lovelock tensor in terms of the Riemann-Christoffel and Ricci curvature tensors and the Riemann curvature scalar for *n*-dimensional differentiable manifolds having a general linear connection. In addition, expressions are given (in the appendix) for the coefficient of the quartic Lovelock Lagrangian as well as for lower-order Lovelock tensors and Lovelock Lagrangian coefficients.

"Since the quartic part of the Lovelock action involves 25 terms, each containing the product of four curvature terms, the writing of the corresponding field equations using the usual method of variation of the Lagrangian with respect to the metric tensor is a formidable task."
— Demaret, J., Y. De Rop, P. Tombal, and A. Moussiaux, "Qualitative Analysis of Ten-dimensional Lovelock Cosmological Models," *Gen. Rel. Grav.*, **24** (1992) 1169.

"Writing explicitly the field equations for general space-times in high dimensions, taking into account all the relevant terms of the Lovelock action, is a very complex task. In the case of, say, ten-dimensional models, it can take weeks to write the equations by hand, and the absolute correctness of the results is not at all guaranteed. As an example of the formidable complexity of this type of calculation, let us note that the quartic part of the Lovelock action involves 25 terms, each of which constituted by the contracted product of four curvature tensors, and, moreover, the resulting field equations, obtained after variation with respect to the metric tensor, are not explicitly known."
— Demaret, J., H. Caprasse, A. Moussiaux, P. Tombal, and D. Papadopoulos, "Ten-dimensional Lovelock-type space-times," *Phys. Rev. D*, **41** (1990) 1163.

"The general expression of the variation of the fourth-order contribution to the Lovelock Lagrangian density has even not yet been obtained."
— *Ibid.*

This letter provides the interested reader with a general expression for the quartic Lovelock tensor $G_{(4)a}{}^{b}$ in terms of the Riemann-Christoffel curvature tensor[1]

$$R_{abc}{}^{d} \equiv 2\,(\partial_{[a}\,\Gamma_{b]}{}^{d}{}_{c} + \Gamma_{[a}{}^{d}{}_{|e|}\,\Gamma_{b]}{}^{e}{}_{c} + \Omega_{a}{}^{e}{}_{b}\,\Gamma_{e}{}^{d}{}_{c}), \tag{1}$$

the Ricci curvature tensor

$$R_{a}{}^{b} \equiv R_{ca}{}^{bc} = -\,R_{ac}{}^{bc} = -\,R_{ca}{}^{cb} + 2\,(\nabla_{[c}\,Q_{a]}{}^{bc} + S_{ca}{}^{d}\,Q_{d}{}^{bc}), \tag{2}$$

and the Riemann curvature scalar

$$R \equiv R_{a}{}^{a} = R_{ba}{}^{ab} = -\,R_{ab}{}^{ab} = R_{ab}{}^{ba} = -\,R_{ba}{}^{ba} \tag{3}$$

using anholonomic coordinates for *n*-dimensional differentiable manifolds having a general linear connection, where $\partial_a$ is the Pfaffian derivative, $\Gamma_{a}{}^{b}{}_{c}$ the connection coefficient, $\Omega_{a}{}^{b}{}_{c}$ the object of anholonomity, $Q_{a}{}^{bc}$ the non-metricity tensor, and $S_{ab}{}^{c}$ the torsion tensor. (An expression for the coefficient $L_{(4)}$ of the quartic Lovelock Lagrangian appears in the appendix.)

Such an expression for $G_{(4)a}{}^{b}$ has evidently not appeared in the open literature hitherto, even though analogues of some terms of $G_{(4)a}{}^{b}$ per Eq. (7) below have already appeared, viz. analogues of the 103*rd*, 104*th*, 107*th*, 109*th*, 111*th*, 112*th*, and 115*th* terms in Grisaru and Zanon[2], of the 109*th* and 111*th* in Freeman *et al.*,[3] and of the 111*th* in Grisaru *et al.*[4]

In accordance with various general definitions given by Müller-Hoissen[5] and Verwimp,[6] $L_{(p)}$ and $G_{(p)a}{}^{b}$ for $p \geq 1$ are given by the formulas

$$L_{(p)} = \frac{(2p)!}{2^p}\,R_{[i_1 i_2}{}^{i_1 i_2}\,R_{i_3 i_4}{}^{i_3 i_4} \ldots R_{i_{2p-1} i_{2p}]}{}^{i_{2p-1} i_{2p}}, \tag{4}$$

and

$$G_{(p)a}{}^{b} = \frac{(2p+1)!}{2^{p+1}p}\,\delta^{b}_{[a}\,R_{i_1 i_2}{}^{i_1 i_2}\,R_{i_3 i_4}{}^{i_3 i_4} \ldots R_{i_{2p-1} i_{2p}]}{}^{i_{2p-1} i_{2p}}, \tag{5}$$

which comprise $(2p)!$ and $(2p+1)!$ tensor index permutations, respectively.

Numerical properties of $L_{(p)}$ and $G_{(p)a}{}^{b}$ for $0 \leq p \leq 4$ are tabulated below. Expressions for $G_{(p)a}{}^{b}$ and $L_{(p)}$ for $0 \leq p \leq 3$ appear in the appendix.

TABLE 1. SOME NUMERICAL PROPERTIES OF $L_{(p)}$ FOR $0 \leq p \leq 4$.

| Order ($p$) | Quantity | Number of Terms | Number of Permutations Comprehended | Sum of Numerical Factors |
|---|---|---|---|---|
| 0 | $L_{(0)}$ | **1** | 1 | 1 |
| 1 | $L_{(1)}$ | **1** | 2 | 1 |
| 2 | $L_{(2)}$ | **3** | 24 | 6 |
| 3 | $L_{(3)}$ | **8** | 720 | 90 |
| 4 | $L_{(4)}$ | **25** | 40,320 | 2520 |

TABLE 2. SOME NUMERICAL PROPERTIES OF $G_{(p)a}{}^{b}$ FOR $0 \leq p \leq 4$.

| Order ($p$) | Quantity | Number of Terms | Number of Permutations Comprehended | Sum of Numerical Factors |
|---|---|---|---|---|
| 0 | $G_{(0)a}{}^{b}$ | **1** | 1 | 1 |
| 1 | $G_{(1)a}{}^{b}$ | **2** | 6 | 1 1/2 |
| 2 | $G_{(2)a}{}^{b}$ | **7** | 120 | 7 1/2 |
| 3 | $G_{(3)a}{}^{b}$ | **26** | 5040 | 105 |
| 4 | $G_{(4)a}{}^{b}$ | **115** | 362,880 | 2835 |

The quartic Lovelock tensor $G_{(4)a}{}^{b}$ is given by the formula

$$G_{(4)a}{}^{b} = \frac{9!}{2^5 \times 4}\,\delta^{b}_{[a}\,R_{i_1 i_2}{}^{i_1 i_2}\,R_{i_3 i_4}{}^{i_3 i_4}\,R_{i_5 i_6}{}^{i_5 i_6}\,R_{i_7 i_8]}{}^{i_7 i_8}, \tag{6}$$

which comprises 362,880 unique covariant index permutations, of which but 115—together with numerical coefficients—suffice for rendering a general expression for $G_{(4)a}{}^{b}$, the final result (after substituting contractions and re-labeling indices) being given by

$$\begin{aligned} G_{(4)a}{}^{b} = \tfrac{1}{8}\,(&\delta^b_a R^4 - 24\,\delta^b_a R^2 R_d{}^c R_c{}^d + 6\,\delta^b_a R^2 R_{ef}{}^{cd} R_{cd}{}^{ef} + 64\,\delta^b_a R\,R_d{}^c R_e{}^d R_c{}^e - 96\,\delta^b_a R\,R_e{}^c R_f{}^d R_{cd}{}^{ef} - 96\,\delta^b_a R\,R_d{}^c R_{fg}{}^{de} R_{ce}{}^{fg} - \\ &- 8\,\delta^b_a R\,R_{ef}{}^{cd} R_{gh}{}^{ef} R_{cd}{}^{gh} + 32\,\delta^b_a R\,R_{eg}{}^{cd} R_{ch}{}^{ef} R_{df}{}^{gh} + 48\,\delta^b_a R_d{}^c R_c{}^d R_f{}^e R_e{}^f - 96\,\delta^b_a R_d{}^c R_e{}^d R_f{}^e R_c{}^f + 384\,\delta^b_a R_d{}^c R_f{}^d R_g{}^e R_{ce}{}^{fg} - \\ &- 24\,\delta^b_a R_d{}^c R_c{}^d R_{gh}{}^{ef} R_{ef}{}^{gh} + 192\,\delta^b_a R_d{}^c R_e{}^d R_{gh}{}^{ef} R_{cf}{}^{gh} + 96\,\delta^b_a R_e{}^c R_f{}^d R_{gh}{}^{ef} R_{cd}{}^{gh} - 192\,\delta^b_a R_e{}^c R_g{}^d R_{ch}{}^{ef} R_{df}{}^{gh} + \\ &+ 192\,\delta^b_a R_e{}^c R_g{}^d R_{dh}{}^{ef} R_{cf}{}^{gh} - 192\,\delta^b_a R_d{}^c R_{cf}{}^{de} R_{hi}{}^{fg} R_{eg}{}^{hi} + 96\,\delta^b_a R_d{}^c R_{fg}{}^{de} R_{hi}{}^{fg} R_{ce}{}^{hi} - 384\,\delta^b_a R_d{}^c R_{fh}{}^{de} R_{ci}{}^{fg} R_{eg}{}^{hi} + \\ &+ 3\,\delta^b_a R_{ef}{}^{cd} R_{cd}{}^{ef} R_{ij}{}^{gh} R_{gh}{}^{ij} - 48\,\delta^b_a R_{ef}{}^{cd} R_{cg}{}^{ef} R_{ij}{}^{gh} R_{dh}{}^{ij} + 6\,\delta^b_a R_{ef}{}^{cd} R_{gh}{}^{ef} R_{ij}{}^{gh} R_{cd}{}^{ij} - 96\,\delta^b_a R_{ef}{}^{cd} R_{gi}{}^{ef} R_{cj}{}^{gh} R_{dh}{}^{ij} + \\ &+ 48\,\delta^b_a R_{eg}{}^{cd} R_{ci}{}^{ef} R_{dj}{}^{gh} R_{fh}{}^{ij} - 96\,\delta^b_a R_{eg}{}^{cd} R_{ci}{}^{ef} R_{fj}{}^{gh} R_{dh}{}^{ij} - 8\,R_a{}^b R^3 + 48\,R_a{}^c R^2 R_c{}^b + 96\,R_a{}^b R\,R_d{}^c R_c{}^d - 192\,R_a{}^c R\,R_d{}^b R_c{}^d + \\ &+ 192\,R_a{}^c R\,R_e{}^d R_{cd}{}^{be} - 24\,R_a{}^b R\,R_{ef}{}^{cd} R_{cd}{}^{ef} + 96\,R_a{}^c R\,R_{ef}{}^{bd} R_{cd}{}^{ef} - 128\,R_a{}^b R_d{}^c R_e{}^d R_c{}^e - 192\,R_a{}^c R_c{}^b R_e{}^d R_d{}^e + 384\,R_a{}^c R_d{}^b R_e{}^d R_c{}^e + \\ &+ 192\,R_a{}^b R_e{}^c R_f{}^d R_{cd}{}^{ef} - 384\,R_a{}^c R_e{}^b R_f{}^d R_{cd}{}^{ef} - 384\,R_a{}^c R_e{}^d R_f{}^e R_{cd}{}^{bf} + 384\,R_a{}^c R_f{}^d R_c{}^e R_{de}{}^{bf} + 192\,R_a{}^b R_d{}^c R_{fg}{}^{de} R_{ce}{}^{fg} + \end{aligned} \tag{7}$$

[1] Schouten, J. A., *Ricci Calculus*, Springer-Verlag, Berlin (1954), p. 171.
[2] Grisaru, M. T., and D. Zanon, "Sigma-Model Superstring Corrections to the Einstein-Hilbert Action," *Phys. Lett.*, **177B** (1986) 347.
[3] Freeman, M. D., C. N. Pope, M. F. Sohnius, and K. S. Stelle, "Higher-Order σ-Model Counterterms and the Effective Action for Superstrings," *Phys. Lett.*, **178B** (1986) 199.
[4] Grisaru, M. T., A. E. M. van de Ven, and D. Zanon, "Four-Loop β-Function for the $N = 1$ and $N = 2$ Supersymmetric Non-Linear Sigma Model in Two Dimensions," *Phys. Lett.*, **173B** (1986) 423.
[5] Müller-Hoissen, F., "Spontaneous Compactification with Quadratic and Cubic Curvature Terms," *Phys. Lett.*, **163B** (1985) 106; "From Chern-Simons to Gauss-Bonnet," *Nucl. Phys.*, **B346** (1990) 235.
[6] Verwimp, T., "On higher dimensional gravity: the Lagrangian, its dimensional reduction and a cosmological model," *Class. Quantum Grav.*, **6** (1989) 1655.

$$
\begin{aligned}
&+ 48\, R_a{}^c R_c{}^b R_{fg}{}^{de} R_{de}{}^{fg} - 192\, R_a{}^c R_d{}^b R_{fg}{}^{de} R_{ce}{}^{fg} - 192\, R_a{}^c R_c{}^d R_{fg}{}^{be} R_{de}{}^{fg} - 192\, R_a{}^c R_e{}^d R_{fg}{}^{be} R_{cd}{}^{fg} + 384\, R_a{}^c R_f{}^d R_{cg}{}^{be} R_{de}{}^{fg} - \\
&- 384\, R_a{}^c R_f{}^d R_{dg}{}^{be} R_{ce}{}^{fg} + 16\, R_a{}^b R_{ef}{}^{cd} R_{gh}{}^{ef} R_{cd}{}^{gh} - 64\, R_a{}^b R_{eg}{}^{cd} R_{ch}{}^{ef} R_{df}{}^{gh} + 192\, R_a{}^c R_{ce}{}^{bd} R_{gh}{}^{ef} R_{df}{}^{gh} - 96\, R_a{}^c R_{ef}{}^{bd} R_{gh}{}^{ef} R_{cd}{}^{gh} - \\
&- 384\, R_a{}^c R_{eg}{}^{bd} R_{dh}{}^{ef} R_{cf}{}^{gh} - 48\, R_{ad}{}^{bc} R^2 R_c{}^d - 24\, R_{ae}{}^{cd} R^2 R_{cd}{}^{be} + 192\, R_{ad}{}^{bc} R\, R_e{}^d R_c{}^e + 192\, R_{ae}{}^{cd} R\, R_c{}^b R_d{}^e - 192\, R_{ae}{}^{bc} R\, R_f{}^d R_{cd}{}^{ef} - \\
&- 96\, R_{ae}{}^{cd} R\, R_f{}^b R_{cd}{}^{ef} + 96\, R_{ae}{}^{cd} R\, R_f{}^e R_{cd}{}^{bf} - 192\, R_{af}{}^{cd} R\, R_c{}^e R_{de}{}^{bf} - 96\, R_{ad}{}^{bc} R\, R_{fg}{}^{de} R_{ce}{}^{fg} + 48\, R_{ae}{}^{cd} R\, R_{fg}{}^{be} R_{cd}{}^{fg} - 192\, R_{af}{}^{cd} R\, R_{cg}{}^{be} R_{de}{}^{fg} + \\
&+ 192\, R_{ad}{}^{bc} R_c{}^d R_f{}^e R_e{}^f - 384\, R_{ad}{}^{bc} R_e{}^d R_f{}^e R_c{}^f - 384\, R_{ae}{}^{cd} R_c{}^b R_f{}^e R_d{}^f - 384\, R_{af}{}^{cd} R_e{}^b R_c{}^e R_d{}^f + 384\, R_{ad}{}^{bc} R_f{}^d R_g{}^e R_{ce}{}^{fg} + \\
&+ 384\, R_{af}{}^{bc} R_c{}^d R_g{}^e R_{de}{}^{fg} + 384\, R_{af}{}^{bc} R_e{}^d R_g{}^e R_{cd}{}^{fg} - 192\, R_{ae}{}^{cd} R_f{}^b R_g{}^e R_{cd}{}^{fg} + 384\, R_{af}{}^{cd} R_c{}^b R_g{}^e R_{de}{}^{fg} + 192\, R_{af}{}^{cd} R_e{}^b R_g{}^e R_{cd}{}^{fg} - \\
&- 384\, R_{af}{}^{cd} R_g{}^b R_c{}^e R_{de}{}^{fg} - 192\, R_{ae}{}^{cd} R_f{}^e R_g{}^f R_{cd}{}^{bg} + 384\, R_{af}{}^{cd} R_c{}^e R_g{}^f R_{de}{}^{bg} - 384\, R_{af}{}^{cd} R_g{}^e R_c{}^f R_{de}{}^{bg} - 192\, R_{ag}{}^{cd} R_c{}^e R_d{}^f R_{ef}{}^{bg} + \\
&+ 384\, R_{ag}{}^{cd} R_f{}^e R_c{}^f R_{de}{}^{bg} + 96\, R_{ag}{}^{cd} R_f{}^e R_e{}^f R_{cd}{}^{bg} - 48\, R_{ad}{}^{bc} R_c{}^d R_{gh}{}^{ef} R_{ef}{}^{gh} + 192\, R_{ad}{}^{bc} R_e{}^d R_{gh}{}^{ef} R_{cf}{}^{gh} + 192\, R_{ae}{}^{bc} R_c{}^d R_{gh}{}^{ef} R_{df}{}^{gh} + \\
&+ 192\, R_{ae}{}^{bc} R_f{}^d R_{gh}{}^{ef} R_{cd}{}^{gh} - 384\, R_{ae}{}^{bc} R_g{}^d R_{ch}{}^{ef} R_{df}{}^{gh} + 384\, R_{ae}{}^{bc} R_g{}^d R_{dh}{}^{ef} R_{cf}{}^{gh} + 192\, R_{ae}{}^{cd} R_c{}^b R_{gh}{}^{ef} R_{df}{}^{gh} + 96\, R_{ae}{}^{cd} R_f{}^b R_{gh}{}^{ef} R_{cd}{}^{gh} + \\
&+ 384\, R_{ag}{}^{cd} R_e{}^b R_{ch}{}^{ef} R_{df}{}^{gh} - 192\, R_{ae}{}^{cd} R_c{}^e R_{gh}{}^{bf} R_{df}{}^{gh} - 96\, R_{ae}{}^{cd} R_f{}^e R_{gh}{}^{bf} R_{cd}{}^{gh} + 384\, R_{ae}{}^{cd} R_g{}^e R_{ch}{}^{bf} R_{df}{}^{gh} + 192\, R_{af}{}^{cd} R_c{}^e R_{gh}{}^{bf} R_{de}{}^{gh} - \\
&- 384\, R_{af}{}^{cd} R_g{}^e R_{ch}{}^{bf} R_{de}{}^{gh} - 192\, R_{af}{}^{cd} R_g{}^e R_{eh}{}^{bf} R_{cd}{}^{gh} - 384\, R_{ag}{}^{cd} R_c{}^e R_{dh}{}^{bf} R_{ef}{}^{gh} + 384\, R_{ag}{}^{cd} R_c{}^e R_{eh}{}^{bf} R_{df}{}^{gh} + 384\, R_{ag}{}^{cd} R_f{}^e R_{ch}{}^{bf} R_{de}{}^{gh} + \\
&+ 192\, R_{ag}{}^{cd} R_f{}^e R_{eh}{}^{bf} R_{cd}{}^{gh} - 192\, R_{ag}{}^{cd} R_h{}^e R_{cd}{}^{bf} R_{ef}{}^{gh} + 384\, R_{ag}{}^{cd} R_h{}^e R_{ce}{}^{bf} R_{df}{}^{gh} - 192\, R_{ad}{}^{bc} R_{cf}{}^{de} R_{hi}{}^{fg} R_{eg}{}^{hi} + 96\, R_{ad}{}^{bc} R_{fg}{}^{de} R_{hi}{}^{fg} R_{ce}{}^{hi} - \\
&- 384\, R_{ad}{}^{bc} R_{fh}{}^{de} R_{ci}{}^{fg} R_{eg}{}^{hi} - 24\, R_{ae}{}^{cd} R_{cd}{}^{be} R_{hi}{}^{fg} R_{fg}{}^{hi} + 192\, R_{ae}{}^{cd} R_{cf}{}^{be} R_{hi}{}^{fg} R_{dg}{}^{hi} - 48\, R_{ae}{}^{cd} R_{fg}{}^{be} R_{hi}{}^{fg} R_{cd}{}^{hi} + 192\, R_{ae}{}^{cd} R_{fh}{}^{be} R_{ci}{}^{fg} R_{dg}{}^{hi} + \\
&+ 96\, R_{af}{}^{cd} R_{cd}{}^{be} R_{hi}{}^{fg} R_{eg}{}^{hi} + 192\, R_{af}{}^{cd} R_{cg}{}^{be} R_{hi}{}^{fg} R_{de}{}^{hi} + 192\, R_{af}{}^{cd} R_{gh}{}^{be} R_{ei}{}^{fg} R_{cd}{}^{hi} + 384\, R_{ah}{}^{cd} R_{cf}{}^{be} R_{di}{}^{fg} R_{eg}{}^{hi} - \\
&- 384\, R_{ah}{}^{cd} R_{cf}{}^{be} R_{ei}{}^{fg} R_{dg}{}^{hi} + 192\, R_{ah}{}^{cd} R_{fg}{}^{be} R_{ci}{}^{fg} R_{de}{}^{hi} + 96\, R_{ah}{}^{cd} R_{fg}{}^{be} R_{ei}{}^{fg} R_{cd}{}^{hi} + 384\, R_{ah}{}^{cd} R_{fi}{}^{be} R_{ce}{}^{fg} R_{dg}{}^{hi}).
\end{aligned}
$$

For a check, note that (1) $G_{(4)a}{}^a = \frac{n-8}{8} L_{(4)}$ and (2) the magnitudes of the numerical coefficients of $G_{(4)a}{}^b$ add up to $\frac{9!}{2^5 \times 4} = \frac{1}{8} \times 22{,}680 = 2835$.

## APPENDIX

The quartic Lovelock Lagrangian coefficient $L_{(4)}$ is given by the formula

$$L_{(4)} = \frac{8!}{2^4} R_{[i_1 i_2}{}^{i_1 i_2} R_{i_3 i_4}{}^{i_3 i_4} R_{i_5 i_6}{}^{i_5 i_6} R_{i_7 i_8]}{}^{i_7 i_8}, \tag{8}$$

which comprises 40,320 unique covariant index permutations, of which but 25—together with numerical coefficients—suffice for rendering a general expression for $L_{(4)}$, the final result (after substituting contractions and re-labeling indices) being given by

$$
\begin{aligned}
L_{(4)} = {}& R^4 - 24\, R^2 R_b{}^a R_a{}^b + 6\, R^2 R_{cd}{}^{ab} R_{ab}{}^{cd} + 64\, R\, R_b{}^a R_c{}^b R_a{}^c - 96\, R\, R_c{}^a R_d{}^b R_{ab}{}^{cd} - 96\, R\, R_b{}^a R_{de}{}^{bc} R_{ac}{}^{de} - 8\, R\, R_{cd}{}^{ab} R_{ef}{}^{cd} R_{ab}{}^{ef} + \\
&+ 32\, R\, R_{ce}{}^{ab} R_{af}{}^{cd} R_{bd}{}^{ef} + 48\, R_b{}^a R_a{}^b R_d{}^c R_c{}^d - 96\, R_b{}^a R_c{}^b R_d{}^c R_a{}^d + 384\, R_b{}^a R_d{}^b R_e{}^c R_{ac}{}^{de} - 24\, R_b{}^a R_a{}^b R_{ef}{}^{cd} R_{cd}{}^{ef} + \\
&+ 192\, R_b{}^a R_c{}^b R_{ef}{}^{cd} R_{ad}{}^{ef} + 96\, R_c{}^a R_d{}^b R_{ef}{}^{cd} R_{ab}{}^{ef} - 192\, R_c{}^a R_e{}^b R_{af}{}^{cd} R_{bd}{}^{ef} + 192\, R_c{}^a R_e{}^b R_{bf}{}^{cd} R_{ad}{}^{ef} - 192\, R_b{}^a R_{ad}{}^{bc} R_{fg}{}^{de} R_{ce}{}^{fg} + \\
&+ 96\, R_b{}^a R_{de}{}^{bc} R_{fg}{}^{de} R_{ac}{}^{fg} - 384\, R_b{}^a R_{df}{}^{bc} R_{ag}{}^{de} R_{ce}{}^{fg} + 3\, R_{cd}{}^{ab} R_{ab}{}^{cd} R_{gh}{}^{ef} R_{ef}{}^{gh} - 48\, R_{cd}{}^{ab} R_{ae}{}^{cd} R_{gh}{}^{ef} R_{bf}{}^{gh} + \\
&+ 6\, R_{cd}{}^{ab} R_{ef}{}^{cd} R_{gh}{}^{ef} R_{ab}{}^{gh} - 96\, R_{cd}{}^{ab} R_{eg}{}^{cd} R_{ah}{}^{ef} R_{bf}{}^{gh} + 48\, R_{ce}{}^{ab} R_{ag}{}^{cd} R_{bh}{}^{ef} R_{df}{}^{gh} - 96\, R_{ce}{}^{ab} R_{ag}{}^{cd} R_{dh}{}^{ef} R_{bf}{}^{gh}.
\end{aligned} \tag{9}
$$

For a check, note that the magnitudes of the numerical coefficients of $L_{(4)}$ add up to $\frac{8!}{2^4} = 2520$.

The Lovelock tensors $G_{(p)a}{}^b$ for $0 \le p \le 3$ are given by

$$G_{(0)a}{}^b = \frac{1!}{2^0} \delta_a^b, \tag{10}$$

$$G_{(1)a}{}^b = \frac{3!}{2^2 \times 1} \delta_{[a}^b R_{i_1 i_2]}{}^{i_1 i_2}, \tag{11}$$

$$G_{(2)a}{}^b = \frac{5!}{2^3 \times 2} \delta_{[a}^b R_{i_1 i_2}{}^{i_1 i_2} R_{i_3 i_4]}{}^{i_3 i_4}, \tag{12}$$

and

$$G_{(3)a}{}^b = \frac{7!}{2^4 \times 3} \delta_{[a}^b R_{i_1 i_2}{}^{i_1 i_2} R_{i_3 i_4}{}^{i_3 i_4} R_{i_5 i_6]}{}^{i_5 i_6}, \tag{13}$$

whence

$$G_{(0)a}{}^b = \delta_a^b, \tag{14}$$

$$G_{(1)a}{}^b = \frac{1}{2} (- \delta_a^b R + 2\, R_a{}^b), \tag{15}$$

$$
\begin{aligned}
G_{(2)a}{}^b = \frac{1}{4} (&\delta_a^b R^2 - 4\, \delta_a^b R_d{}^c R_c{}^d + \delta_a^b R_{ef}{}^{cd} R_{cd}{}^{ef} - 4\, R_a{}^b R + \\
&+ 8\, R_a{}^c R_c{}^b - 8\, R_{ad}{}^{bc} R_c{}^d - 4\, R_{ae}{}^{cd} R_{cd}{}^{be}),
\end{aligned} \tag{16}
$$

and

$$
\begin{aligned}
G_{(3)a}{}^b = \frac{1}{6} (&- \delta_a^b R^3 + 12\, \delta_a^b R\, R_d{}^c R_c{}^d - 3\, \delta_a^b R\, R_{ef}{}^{cd} R_{cd}{}^{ef} - 16\, \delta_a^b R_d{}^c R_e{}^d R_c{}^e + 24\, \delta_a^b R_e{}^c R_f{}^d R_{cd}{}^{ef} + 24\, \delta_a^b R_d{}^c R_{fg}{}^{de} R_{ce}{}^{fg} + 2\, \delta_a^b R_{ef}{}^{cd} R_{gh}{}^{ef} R_{cd}{}^{gh} - \\
&- 8\, \delta_a^b R_{eg}{}^{cd} R_{ch}{}^{ef} R_{df}{}^{gh} + 6\, R_a{}^b R^2 - 24\, R_a{}^c R\, R_c{}^b - 24\, R_a{}^b R_d{}^c R_c{}^d + 48\, R_a{}^c R_d{}^b R_c{}^d - 48\, R_a{}^c R_e{}^d R_{cd}{}^{be} + 6\, R_a{}^b R_{ef}{}^{cd} R_{cd}{}^{ef} - \\
&- 24\, R_a{}^c R_{ef}{}^{bd} R_{cd}{}^{ef} + 24\, R_{ad}{}^{bc} R\, R_c{}^d + 12\, R_{ae}{}^{cd} R\, R_{cd}{}^{be} - 48\, R_{ad}{}^{bc} R_e{}^d R_c{}^e - 48\, R_{ae}{}^{cd} R_c{}^b R_d{}^e + 48\, R_{ae}{}^{bc} R_f{}^d R_{cd}{}^{ef} + 24\, R_{ae}{}^{cd} R_f{}^b R_{cd}{}^{ef} - \\
&- 24\, R_{ae}{}^{cd} R_f{}^e R_{cd}{}^{bf} + 48\, R_{af}{}^{cd} R_c{}^e R_{de}{}^{bf} + 24\, R_{ad}{}^{bc} R_{fg}{}^{de} R_{ce}{}^{fg} - 12\, R_{ae}{}^{cd} R_{fg}{}^{be} R_{cd}{}^{fg} + 48\, R_{af}{}^{cd} R_{cg}{}^{be} R_{de}{}^{fg}).
\end{aligned} \tag{17}
$$

For a check, note—for $1 \le p \le 3$—that (1) $G_{(p)a}{}^a = \frac{n-2p}{2p} L_{(p)}$ and (2) the magnitudes of the numerical coefficients of $G_{(p)a}{}^b$ add up to $\frac{(2p+1)!}{2^{p+1} p}$.

The Lovelock Lagrangian coefficients $L_{(p)}$ for $0 \le p \le 3$ are given by

$$L_{(0)} = \frac{0!}{2^0} = 1, \tag{18}$$

$$L_{(1)} = \frac{2!}{2^1} R_{[i_1 i_2]}{}^{i_1 i_2}, \tag{19}$$

$$L_{(2)} = \frac{4!}{2^2} R_{[i_1 i_2}{}^{i_1 i_2} R_{i_3 i_4]}{}^{i_3 i_4}, \tag{20}$$

and

$$L_{(3)} = \frac{6!}{2^3} R_{[i_1 i_2}{}^{i_1 i_2} R_{i_3 i_4}{}^{i_3 i_4} R_{i_5 i_6]}{}^{i_5 i_6}, \tag{21}$$

whence

$$L_{(0)} = 1, \tag{22}$$

$$L_{(1)} = - R, \tag{23}$$

$$L_{(2)} = R^2 - 4\, R_b{}^a R_a{}^b + R_{cd}{}^{ab} R_{ab}{}^{cd}, \tag{24}$$

and

$$
\begin{aligned}
L_{(3)} = {}& - R^3 + 12\, R\, R_b{}^a R_a{}^b - 3\, R\, R_{cd}{}^{ab} R_{ab}{}^{cd} - 16\, R_b{}^a R_c{}^b R_a{}^c + \\
&+ 24\, R_c{}^a R_d{}^b R_{ab}{}^{cd} + 24\, R_b{}^a R_{de}{}^{bc} R_{ac}{}^{de} + \\
&+ 2\, R_{cd}{}^{ab} R_{ef}{}^{cd} R_{ab}{}^{ef} - 8\, R_{ce}{}^{ab} R_{af}{}^{cd} R_{bd}{}^{ef}.
\end{aligned} \tag{25}
$$

For a check, note that (1) the magnitudes of the numerical coefficients of $L_{(p)}$ add up—for any $p$—to $\frac{(2p)!}{2^p}$ and (2) $\frac{\partial}{\partial R} L_{(p)} = - p\, L_{(p-1)}$ for $p \ge 1$.